\documentstyle[e-e-ijmpa,twoside,citepunct,epsfig]{article}

\renewcommand{\thefootnote}{\fnsymbol{footnote}}        

\begin{document}

\normalsize\textlineskip
\pagestyle{empty}

\title{\bf PRECISION MASS DETERMINATION OF THE HIGGS BOSON \\
AT PHOTON-PHOTON COLLIDERS%
\footnote{This work was supported in part by the U.S. Department of
Energy under Contract No.~DE-AC03-76SF00098.}
}

\author{TOMOMI OHGAKI}

\address{Lawrence Berkeley National Laboratory \\
         Berkeley, California 94720, USA}

\maketitle\abstracts{We demonstrate a measurement of the Higgs boson
mass by the method of energy scanning at photon-photon colliders,
using the high energy edge of the photon spectrum. With an integrated
luminosity of 50~$\rm{fb^{-1}}$ it is possible to measure the standard
model Higgs mass to within 110 MeV in photon-photon collisions for
$m_h=100$ GeV. As for the total width of the Higgs boson, the
statistical error $\Delta\Gamma_h/\Gamma_{h~\rm{SM}}=0.06$ is expected
for $m_h=100$ GeV, if both $\Gamma(h\to\gamma\gamma)$ and $\Gamma(h\to
b\bar{b})$ are fixed at the predicted standard model value.}

\setcounter{footnote}{0}
\renewcommand{\thefootnote}{\alph{footnote}}


\vspace*{20pt}\textlineskip      
\section{Introduction}          
\begin{figure}[htbp]
\vspace*{13pt}
\centerline{\epsfig{file=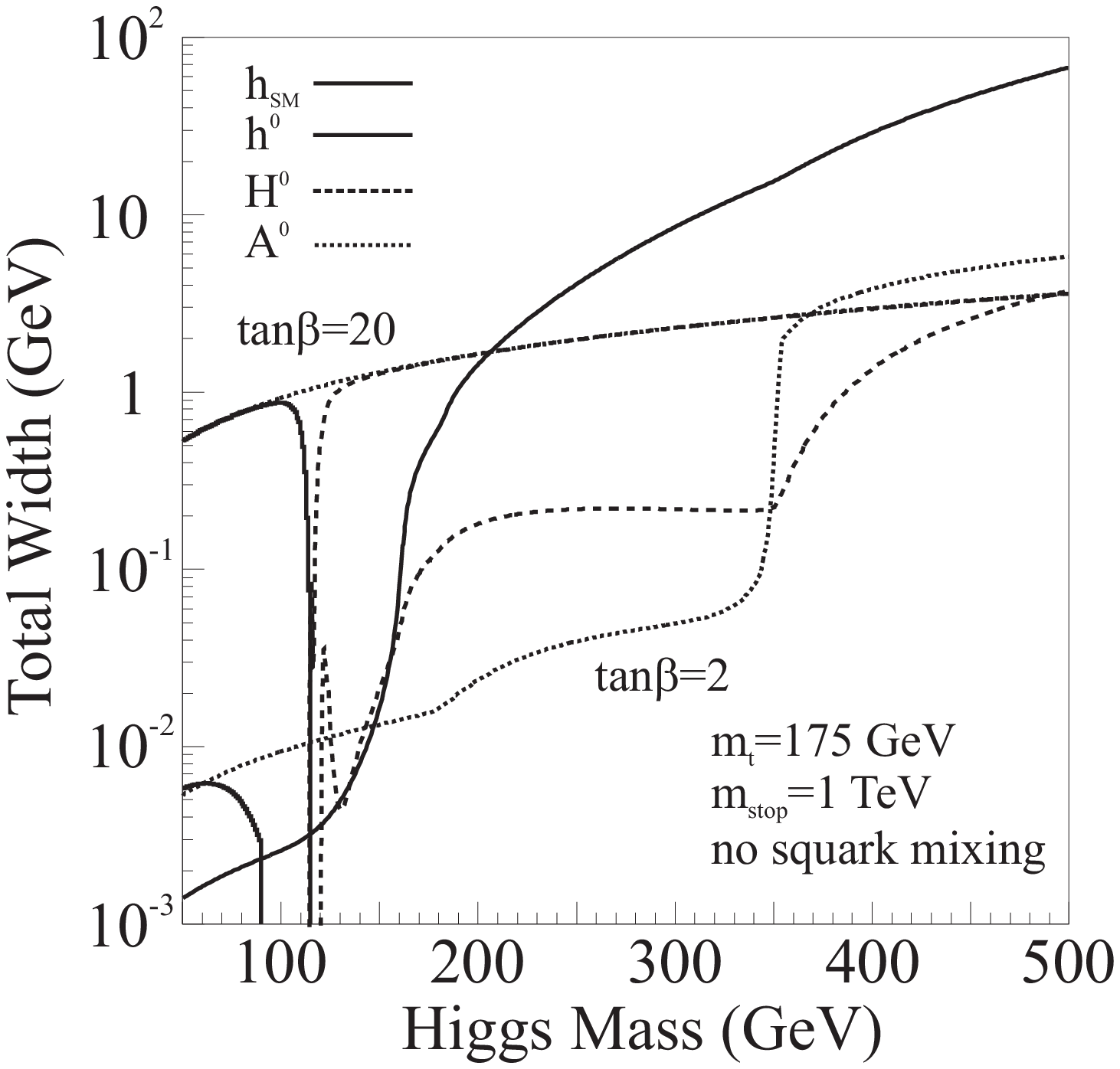,width=6.1cm}}
\vspace*{13pt}
\fcaption{The total widths of the SM and MSSM Higgs bosons. The top
quark mass is assumed to be 175 GeV. In the case of the MSSM, the
results for $\tan\beta=2$ and 20 are shown, taking $m_{\tilde{t}}=1$
TeV, including two-loop radiative corrections, and neglecting squark
mixing. SUSY decay channels are assumed to be absent. Computed by
HDECAY~\protect\cite{djo98}.}
\label{fig:higgs}
\end{figure}

\vspace*{-0.5pt}
  One of the most important tasks of the current and future collider
experiments will be to detect and study Higgs boson(s). The accuracy
of the measurement of the Higgs boson mass will impact precision tests
of loop corrections, both in the standard model (SM) and in the
extended models such as the minimal supersymmetric model
(MSSM)~\cite{gun97,bar971,bar972}. Deviations of the total widths of
the Higgs bosons from SM predictions can be directly compared to
predictions of alternative models such as the MSSM, the non-minimal
supersymmetric standard model, or the general two-Higgs-doublet
model~\cite{gun97,bar971,bar972}. The total widths for the SM Higgs
boson $h_{SM}$ and the three neutral Higgs bosons $h^0,H^0,A^0$ of the
MSSM are shown in Fig.~\ref{fig:higgs}.

  The interaction of high energy photons at a photon-photon
collider~\cite{gin83,tel95,tel98} provides us with an unique
opportunity to study Higgs boson, because the SM Higgs boson in 
$s$-channel resonance can be produced at photon-photon
colliders~\cite{bor92,bor93,bor94,jik96,ohg97}. In this paper we
point out precision measurements of mass ($m_h$) and total width
($\Gamma_h$) of the Higgs boson by the method of energy scanning,
using the high energy edge of the photon spectrum.

  The method of energy scanning at photon-photon colliders was
first mentioned by V.~Telnov~\cite{tel98}. The luminosity of the
photon-photon collider has a very sharp edge at high energy, much
narrower than the width of the luminosity peak. If the Higgs boson is
a very narrow resonance, we will observe a rapid increase in the
visible cross section of the Higgs production during energy
scanning. 

\newpage
\section{Luminosity of Photon-Photon Colliders}            
\begin{figure}[htbp]
\vspace*{13pt}
\centerline{\epsfig{file=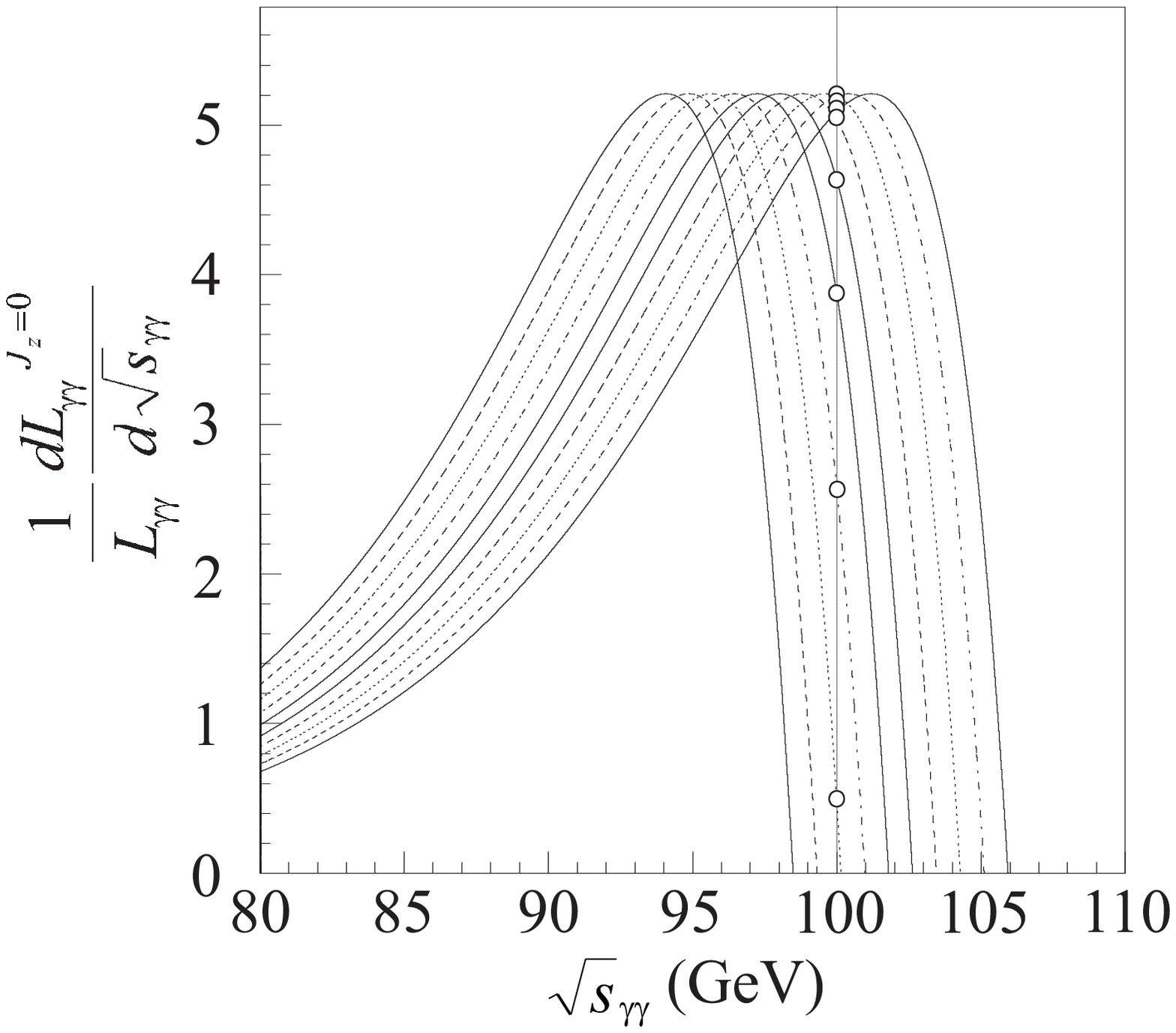,width=6.1cm}}
\vspace*{13pt}
\fcaption{Ten differential luminosities with $J_z=0$ as a function of
the center-of-mass energy in a photon-photon collider for energy
scanning at $m_h=100$ GeV.}
\label{fig:lum}
\end{figure}

  Figure~\ref{fig:lum} shows ten differential luminosities with the
$J_z=0$ angular momentum state of initial photon collisions in a
photon-photon collider for energy scanning at $m_h=100$ GeV. In this
study, we have scanned the Higgs boson resonance from the left side to
the right side in Fig.~\ref{fig:lum}. The circles exhibit the
luminosity points in contact with the Higgs boson and the rise of the
luminosity at $m_h=100$ GeV is rapid at the threshold of energy
scanning. Here we introduce the required parameters for the luminosity
calculation. A laser photon of energy $\omega_L$ is scattered by an
electron beam of energy $E_e$ in the conversion region of the
photon-photon collider. The kinematics of Compton scattering is
characterized by the dimensionless parameter~\cite{gin83}
\begin{eqnarray}
\label{eq:1}
  x\equiv\frac{4E_e\omega_L}{m_e^2}\approx15.3\left[\frac{E_e}{\rm{TeV}}\right]\left[\frac{\omega_L}{\rm{eV}}\right],
\end{eqnarray}
where $m_e$ is electron mass. The maximum energy of the scattered
photon $\omega_{max}$ is $E_e x/(x+1)$ given by $x$. The parameter $x$
is fixed to be 4.8, and we get $\omega_{max}=100$ GeV when $E_e=121$
GeV and $\omega_L=2.6$ eV. The combination of the polarizations of the
electron $P_e$ and the laser $P_L$ should be $P_L P_e=-1$ so that the
generated photon spectrum peaks at its maximum energy. 

  The differential luminosity distribution depends on the variable
$\rho=b/(\gamma a)$, where $a$ is the rms radius of the electron beam
at the interaction point (IP), $b$ is the distance between the
conversion point (CP) and the IP, and $\gamma=E_e/m_e$. The polarized
luminosities with the $J_z=0$ and the $J_z=\pm2$ in a photon-photon
collider were used in Ref.~\citenum{gin83}. Here we assumed $\rho=1$
and the conversion coefficient $k=0.6$. It should be noted that the
shape of the high energy edge and $w_{max}$ are influenced by
nonlinear effects due to very strong focus of the laser field at the
CP. Prior to the actual energy scan, we need to have a fairly good
estimate for nonlinear effects including the polarization.


\section{Higgs Boson and Backgrounds}            
  Once the Higgs boson is observed at future $e^+e^-$ colliders, we
must determine its precise mass and width in order to reveal the
fundamental properties of the Higgs boson. At a photon-photon
collider, the feasibility of the measurement of the two-photon decay
width of a Higgs boson has been studied in the mass range
$M_W<m_h<2M_W$~\cite{bor92,bor93,bor94,jik96,ohg97}. For $m_h<2M_W$,
the SM Higgs boson mainly decays into a $b\bar{b}$ pair and the
daughter $b$-flavored hadrons will be easily identified due to their
long lifetime; therefore, the $b\bar{b}$ events are the best
signals. The cross section of the Higgs boson production can be
described by the Breit-Wigner approximation:
\begin{eqnarray}
\label{eq:4}
  \sigma_{\gamma\gamma\to h\to b\bar{b}}(\sqrt{s})=8\pi\frac{\Gamma(h\to\gamma\gamma)\Gamma(h\to b\bar{b})}{(s-m_h^2)^2+m_h^2\Gamma_h^2}(1+\lambda_1\lambda_2),
\end{eqnarray}
where $\Gamma(h\to\gamma\gamma)$ and $\Gamma(h\to b\bar{b})$ are the decay
widths of the Higgs boson into two photons and a $b\bar{b}$ pair,
$\lambda_1$ and $\lambda_2$ the initial photon helicities,
respectively. The effective cross section of the signal events within
$m_h-\delta<\sqrt{s}<m_h+\delta$ is 
\begin{eqnarray}
\label{eq:5}
  \sigma^{\rm{eff}}_{\gamma\gamma\to h\to
  b\bar{b}}=\int_{m_h-\delta}^{m_h+\delta}16\pi\frac{\Gamma(h\to\gamma\gamma)\Gamma(h\to
  b\bar{b})}{(\hat{s}-m_h^2)^2+m_h^2\Gamma_h^2}\frac{1}{L_{\gamma\gamma}}\frac{dL_{\gamma\gamma}^{J_z=0}}{d\sqrt{\hat{s}}}d\sqrt{\hat{s}},
\end{eqnarray}
where $\delta$ expresses the effect of the detector resolution and we
assumed $\delta=5$ GeV. Here we supposed that the total luminosity is
$L_{\gamma\gamma}=L_{\gamma\gamma}^{J_z=0}+L_{\gamma\gamma}^{J_z=\pm2}$.

  The main background processes may be the continuum $\gamma\gamma\to
b\bar{b}$, $c\bar{c}$ as well as the radiative processes
$\gamma\gamma\to b\bar{b}g$, $c\bar{c}g$. The continuum backgrounds
dominantly produced by initial photon collisions in the $J_z=\pm2$ can
be suppressed by controlling the polarization of the colliding photon
beams. Several authors reported that the effect of QCD corrections to
$\gamma\gamma\to q\bar{q}$ is large since the helicity suppression
which affects the background $q\bar{q}$ events does not work due to a
gluon emission~\cite{bor94,jik96}. Recently leading double-logarithmic
QCD corrections for $J_z=0$ were resummed to all orders and the
account of non-Sudakov form factor to higher orders makes the
cross-section well defined and positive definite in all regions of the
phase space~\cite{mel99}. In this study we take account of the
one-loop QCD corrections of the soft gluon emission, hard gluon
emission, and virtual correction, where higher order double
logarithmic corrections are not taken into account~\cite{jik96}. The
effective cross section of the background process $\gamma\gamma\to
b\bar{b}(g)$ or $c\bar{c}(g)$ within $m_h-\delta<\sqrt{s}<m_h+\delta$
is
\begin{eqnarray}
\label{eq:6}
  \sigma^{\rm{eff}}_{\rm{bg}}=\int_{m_h-\delta}^{m_h+\delta}\sigma_{\rm{bg}}(\sqrt{\hat{s}})\frac{1}{L_{\gamma\gamma}}\frac{dL_{\gamma\gamma}}{d\sqrt{\hat{s}}}d\sqrt{\hat{s}},
\end{eqnarray}
where $\sigma_{\rm{bg}}(\sqrt{s})$ is the cross section of the
background process. 

  Since the cross section of $\gamma\gamma\to c\bar{c}$ is larger than
that of $\gamma\gamma\to b\bar{b}$ due to the large electric charge of
the quark, we apply the $b$ tagging in order to eliminate the charm
and the light quark backgrounds. By the topological vertexing
method~\cite{jac97} and the LC Vertex Detector design~\cite{bur99},
the efficiency and purity of $b$-quark jet identification are 70\% and
99\%, respectively. Therefore the tagging efficiencies of
$b\bar{b}(g)$ and $c\bar{c}(g)$ events are assumed as 49\% and 0.005\%
with double tagging, respectively.\footnote{The interaction region
(IR) at photon-photon colliders is complicated, because there are the
sweeping magnet for spent electrons and the optical mirror system for
laser focusing around the vertex detector. We need to study the
performance of the vertex detector at the IR.} \ \cite{wat98} We
impose the following cuts to remove backgrounds: (1) the double
$b\bar{b}$ tagging in the event; (2) $|\cos\theta_{b,\bar{b}}|<0.95$,
where $\theta_{b,\bar{b}}$ is the scattering angle of the $b(\bar{b})$
quark; (3) $|M_{b\bar{b}}-m_h|<5$ GeV.


\section{Results and Discussion}            
\begin{figure}[htbp]
\begin{minipage}[t]{6.3cm}
\vspace*{13pt}
\centerline{\epsfig{file=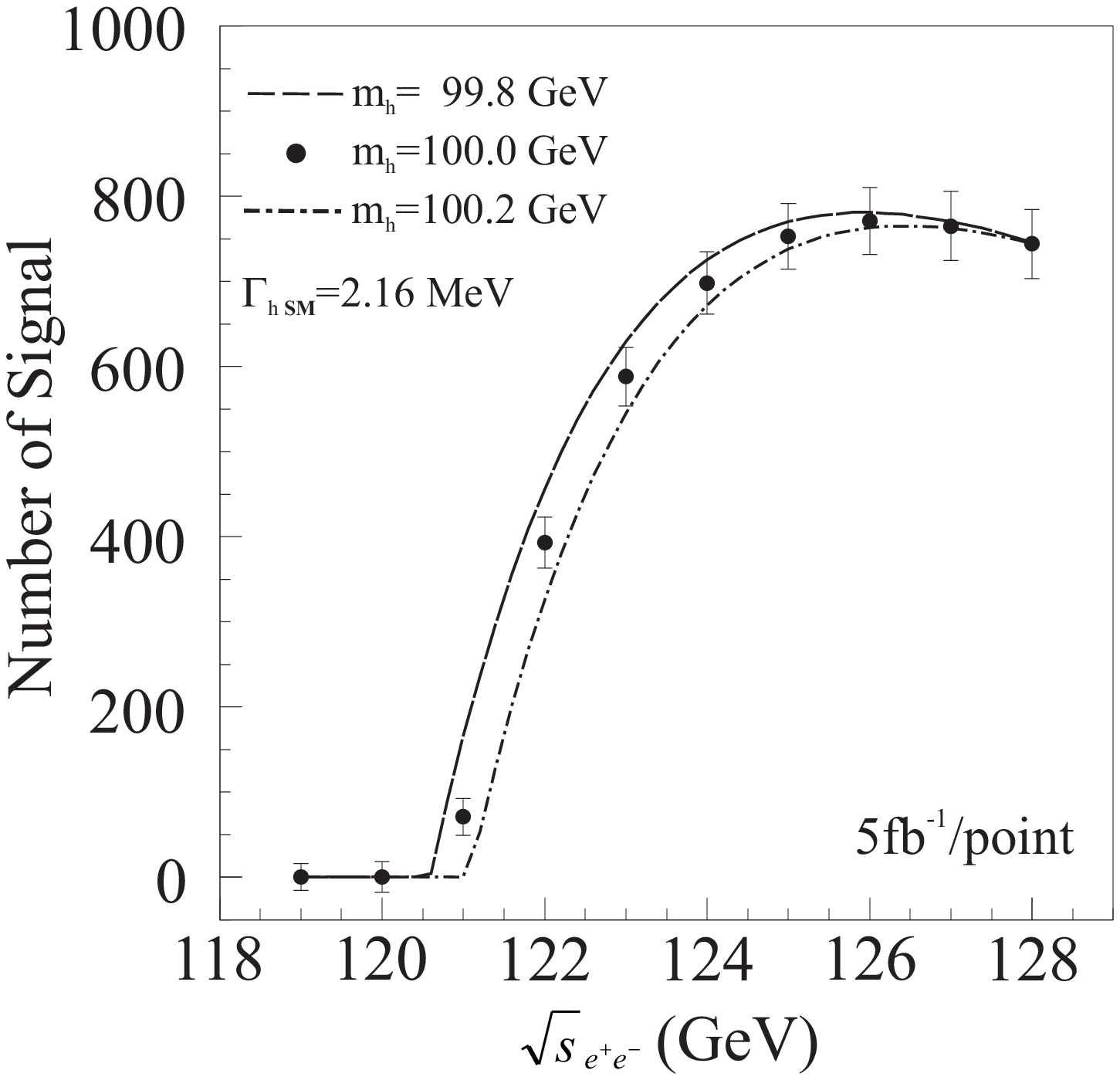,width=6.09cm}}
\vspace*{13pt}
\fcaption{An example of energy scan to determine $m_h$ where each
point corresponds to 5~$\rm{fb}^{-1}$.}
\label{fig:mass}
\end{minipage}
\begin{minipage}[t]{6.3cm}
\vspace*{13pt}
\centerline{\epsfig{file=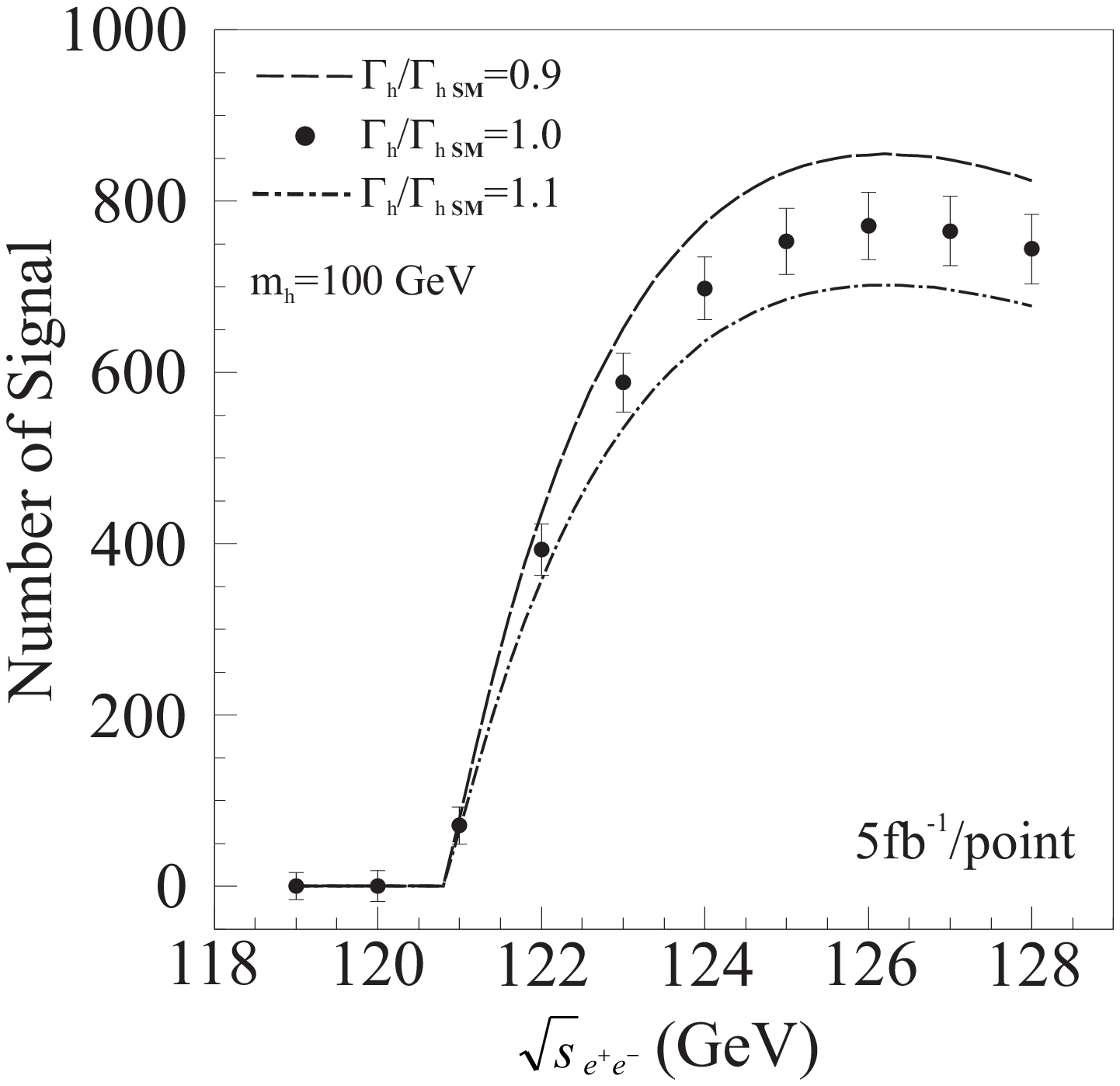,width=6.09cm}}
\vspace*{13pt}
\fcaption{An example of energy scan to determine $\Gamma_h$ where each
point corresponds to 5~$\rm{fb}^{-1}$.}
\label{fig:width}
\end{minipage}
\end{figure}

  Figure~\ref{fig:mass} shows an example of energy scan to determine 
$m_h$. Each energy point corresponds to 5 $\rm{fb}^{-1}$ and the total 
luminosity of photon-photon collisions is 50 $\rm{fb}^{-1}$ in the
same distributions as with Fig.~\ref{fig:lum}. The total width of the
SM Higgs boson $\Gamma_{h~\rm{SM}}$ for $m_{h}$=100 GeV is 2.16 MeV,
which is computed by the HDECAY program~\cite{djo98}. The partial
widths $\Gamma(h\to\gamma\gamma)$ and $\Gamma(h\to b\bar{b})$ at the
predicted SM value with $m_h=$ 100 GeV are fixed for energy scanning at
$m_h=$ 99.8, 100, 100.2 GeV. The statistical errors in
Fig.~\ref{fig:mass} indicate $\sqrt{S+B}$, where $S$ and $B$ are the
numbers for signal and background events. From Fig.~\ref{fig:mass}, we
can understand that the method of energy scanning for $m_h$ is more
effective than that of the measurement of a single point at the
luminosity peak using the same total luminosity, because the
statistical errors at the threshold of energy scanning are smaller
than that at the luminosity peak and we can distinguish the mass
difference of 200 MeV. With the energy scanning of 10 points, the
attainable error in $m_h$ is about 110 MeV at the $1\sigma$ level.

  The measurement for the determination of $\Gamma_h$ by the method of
energy scan is shown in Fig.~\ref{fig:width}. Each energy point
corresponds to 5 $\rm{fb}^{-1}$ and the total luminosity is 50
$\rm{fb}^{-1}$. The partial widths $\Gamma(h\to\gamma\gamma)$ and
$\Gamma(h\to b\bar{b})$ at the predicted SM value with $m_h=100$ GeV are
fixed for energy scanning $\Gamma_h/\Gamma_{h~\rm{SM}}=$ 0.9, 1.0,
1.1. The large difference between the total widths at the luminosity
peak can be seen easily in Fig.~\ref{fig:width}. The statistical error
in $\Gamma_h$ is about 6\% at the $1\sigma$ level. If there are
additional invisible decay modes of Higgs boson, only the total decay
width increases keeping the partial widths of two photons and a
$b\bar{b}$ pair unchanged. In this study we find
$\Gamma_h/\Gamma_{h_{\rm{SM}}}>1$. Of course, this deviation from the
SM should have also been observed in the parent $e^+e^-$
collider. However, this will be independent observation in gamma-gamma
energy scan, which confirms the $e^+e^-$ result.

  Here we consider two cases for the photon-photon collider. First, we
choose $x=4.8$ while tuning the energies of the laser photon and the
electron beam while tuning the scan. Second, we fix the laser energy
and only the energy of the electron beam is tuned during this
scan. The two cases are called the tunable and fixed cases,
respectively.
\begin{table}[htbp]
\tcaption{The statistical errors of the SM Higgs boson mass by
devoting 50/10 $\rm{fb}^{-1}$ to each point. The results in the
parentheses are calculated with the tagging efficiencies 70\% and
3.5\% of $b\bar{b}(g)$ and $c\bar{c}(g)$ events, respectively.}
\label{tbl:mass}
\centerline{\footnotesize\smalllineskip
\begin{tabular}{lcccccc}
\hline
                       & & & $\Delta m_{h_{\rm{SM}}}$ & (MeV) & & \\
  $m_{h_{\rm{SM}}}$ (GeV) &  80 &  90 & 100 & 110 & 120 & 140 \\
\hline
  Tunable case            & +140 & +120 & +100 (+140) & +140 & +170 & +210 \\
                          & $-150$ & $-140$ & $-110$ $(-160)$ & $-100$
  & $-100$ & $-220$ \\
\hline
  Fixed case              & +140 & +120 & +100 (+190) & +170 & +200 & +220 \\
                          & $-170$ & $-130$ & $-120$ $(-160)$ & $-110$
  & $-130$ & $-270$ \\
\hline\\
\end{tabular}}
\end{table}
\begin{table}[htbp]
\tcaption{The statistical errors of the total width of the SM Higgs
boson by devoting 50/10 $\rm{fb}^{-1}$ to each point. The results in
the parentheses are calculated with the tagging efficiencies 70\%
and 3.5\% of $b\bar{b}(g)$ and $c\bar{c}(g)$ events, respectively.}
\label{tbl:width}
\centerline{\footnotesize\smalllineskip
\begin{tabular}{lcccccc}\\
\hline
                       & & & $\Delta \Gamma_h / \Gamma_{h_{\rm{SM}}}$
  & (\%) & & \\
  $m_{h_{\rm{SM}}}$ (GeV) & 80 & 90 & 100 & 110 & 120 & 140 \\
\hline
  Tunable case  & +7.8 & +6.7 & +6.0 (+7.8) & +5.7 & +5.7 & +7.5 \\
                & $-7.0$ & $-6.1$ & $-5.6$ $(-6.9)$ & $-5.3$ & $-5.6$ & $-6.8$
  \\
\hline
  Fixed case    & +8.5 & +7.3 & +6.6 (+8.5) & +6.2 & +6.3 & +8.1 \\
                & $-7.6$ & $-6.6$ & $-6.0$ $(-7.5)$ & $-5.7$ & $-5.8$
  & $-7.3$ \\
\hline\\
\end{tabular}}
\end{table}

  Table~\ref{tbl:mass} lists the statistical errors of the SM Higgs
boson mass at the $1\sigma$ level, using an integrated luminosity of
50/10 $\rm{fb}^{-1}$. In this table, the mass errors of the tunable
case are almost smaller than those of the fixed case. Since the
background processes $\gamma\gamma\to q\bar{q}(g)$ are increasing at
the lower Higgs mass and the branching ratio $B(h\to b\bar{b})$ is
decreasing at the higher Higgs mass, the errors of the Higgs boson
mass near 100 GeV are the smallest. The statistical errors
$\sqrt{S+B}/S$ of the total width $\Gamma_h/\Gamma_{h~\rm{SM}}$ of the
SM Higgs boson with a 50 $\rm{fb}^{-1}$ luminosity are listed in
Table~\ref{tbl:width}. The statistical errors of the total width for
intermediate-mass Higgs bosons are almost within 8\% in
Table~\ref{tbl:width}. Comparatively the results with the tagging
efficiencies 70\% and 3.5\% of $b\bar{b}(g)$ and $c\bar{c}(g)$ events
are listed in Tables~\ref{tbl:mass} and~\ref{tbl:width}.
\begin{table}[htbp]
\tcaption{The expected precision for the mass of the Higgs boson with
$m_{h_{\rm{SM}}}=100$ GeV at the future
colliders~\protect\cite{gun97,bar971}. The NLC threshold result is at
$\sqrt{s}=m_Z+m_{h_{\rm{SM}}}+0.5$ GeV including the initial state
radiation and the beam energy spread~\protect\cite{bar971}. The LHC
error is for ATLAS+CMS~\protect\cite{gun97}. The error at the muon
collider is devoted to the scan with beam energy resolution of
$0.01\%$~\protect\cite{gun97}.}
\label{tbl:coll}
\centerline{\footnotesize\smalllineskip
\begin{tabular}{lccccc}\\
\hline
      & NLC (threshold) & LHC & Muon Collider & Photon-Photon Collider \\
\hline
  $\Delta m_{h_{\rm{SM}}}$ (MeV) & 60 & 95 & 0.1 & 110 (90) \\
  Luminosity ($\rm{fb}^{-1}$) & 100 & 600 & 200 & 50 (100) \\
\hline\\
\end{tabular}}
\end{table}

  At the future colliders, the expected precision for the mass of the
Higgs boson with $m_{h_{\rm{SM}}}$=100 GeV is listed in
Table~\ref{tbl:coll}. The NLC threshold result is at
$\sqrt{s}=m_Z+m_{h_{\rm{SM}}}+0.5$ GeV including the initial state
radiation and the beam energy spread~\cite{bar971}. The LHC error is
for ATLAS+CMS including the statistical and systematic
errors~\cite{gun97}. The error at the muon collider is devoted to the
scan with beam energy resolution of $0.01\%$~\cite{gun97}. From the
table, the accuracy of the Higgs boson mass at the muon collider is
the highest, however the systematic error at the muon collider is
neglected assuming accurate beam energy determination. The accuracy at
the photon-photon collider is 1.5 times lower than that at the NLC
threshold case. Therefore we can perform the complementary measurement
of Higgs boson mass at photon-photon colliders.

  As for other origins of the errors, we need to know the systematic 
uncertainties on the luminosity distribution. The possibilities of the
luminosity measurements at photon-photon colliders have been studied
using the process $\gamma\gamma\to l^+l^-$ or $\gamma\gamma\to
W^+W^-$~\cite{wat98,wat93}. For energy scanning the measurement of the
luminosity distribution at the high energy-edge is crucial and we need
to study it further.


\section{Summary}            
  In this paper, we have shown that it is possible to determine the
Higgs boson mass to a high precision by the method of energy scanning
at photon-photon colliders, using the high energy edge of the photon
spectrum.


\nonumsection{Acknowledgments}
  I express sincere thanks to T.~Takahashi, T.~Tauchi, V.~Telnov, 
I.~Watanabe, M.~Xie and K.~Yokoya for useful discussions.


\nonumsection{References}


\begin{thebibliography}{000}

\bibitem{gun97} J.F.~Gunion {\it et al.}, in {\it Proceedings of the
1996 DPF/DPB Summer Study on New Directions for High-Energy Physics
(Snowmass, 96)}, Snowmass, CO, June 25-July 12, 1996, p.541.
 
\bibitem{bar971} V.~Barger, M.S.~Berger, J.F.~Gunion, and T.~Han,
Phys. Rev. Lett. {\bf78}, 3991 (1997).

\bibitem{bar972} V.~Barger, M.S.~Berger, J.F.~Gunion, and T.~Han,
Phys. Rept. {\bf286}, 1 (1997).

\bibitem{gin83} I.F.~Ginzburg, G.L.~Kotkin, V.G.~Serbo, and
V.I.~Telnov, Nucl. Instrum. and Methods {\bf205}, 47 (1983); A
{\bf219}, 5 (1984). 

\bibitem{tel95} V.~Telnov, Nucl. Instrum. and Methods Phys. Res. A
{\bf355}, 3 (1995).

\bibitem{tel98} V.~Telnov, in {\it Proceedings of the 2nd
International Workshop on Electron-Electron Interactions at TeV
Energies}, Santa Cruz, CA, Sep 22-24, 1997, Int. J. Mod. Phys. A
{\bf13}, 2399 (1998).

\bibitem{bor92} D.L.~Borden, D.A.~Bauer, and D.O.~Caldwell,
SLAC-PUB-5715, 1992.

\bibitem{bor93} D.L.~Borden, D.A.~Bauer, and D.O.~Caldwell,
Phys. Rev. D {\bf48}, 4018 (1993).

\bibitem{bor94} D.L.~Borden, V.A.~Khoze, W.J.~Stirling, and
J.~Ohnemus, Phys. Rev. D {\bf50}, 4499 (1994).

\bibitem{jik96} G.~Jikia and A.~Tkabladze, Phys. Rev. D {\bf54}, 2030
(1996).

\bibitem{ohg97} T.~Ohgaki, T.~Takahashi, and I.~Watanabe, Phys. Rev. D
{\bf56}, 1723 (1997).

\bibitem{mel99} M.~Melles and W.J.Stirling, Phys. Rev. D {\bf59},
094009 (1999); Eur. Phys. J C {\bf9}, 101 (1999); DTP-98-100 (1998);
M.~Melles, W.J.~Stirling, and V.A.~Khoze, DTP-99-70 (1999).

\bibitem{jac97} D.J.~Jackson, Nucl. Instrum. Meth. A {\bf388} 247
(1997).

\bibitem{bur99} LCFI Collaboration, P.N.~Burrows {\it et al.}, in {\it
Proceedings of the 8th International Workshop on Vertex Detectors
(Vertex 99)}, Texel, Netherlands, June 20-25, 1999.

\bibitem{wat98} I.~Watanabe {\it et al.}, Report No. KEK-97-17, 1998.

\bibitem{djo98} A.~Djouadi, J.~Kalinowski, and M.~Spira,
Comput. Phys. Commun. {\bf108}, 56 (1998).

\bibitem{wat93} Y.~Yasui, I.~Watanabe, J.~Kodaira, and I.~Endo,
Nucl. Instrum. and Methods Phys. Res. A {\bf335}, 385 (1993).

\end{thebibliography}
\end{document}